\documentclass[prl,twocolumn,showpacs,superscriptaddress]{revtex4}

\usepackage{graphicx}
\usepackage{dcolumn}
\usepackage{amsmath}
\usepackage{color}

\newcommand{\SrCoVO}{SrCo$_2$V$_2$O$_8$}
\newcommand{\BaCoVO}{BaCo$_2$V$_2$O$_8$}

\begin{document}

\title{Spinon confinement in the one-dimensional Ising-like antiferromagnet \SrCoVO}

\author{Zhe~Wang}
\email{zhe.wang@physik.uni-augsburg.de}
\affiliation{Experimental Physics V, Center for Electronic
Correlations and Magnetism, Institute of Physics, University of Augsburg, 86135 Augsburg, Germany}

\author{M.~Schmidt}
\affiliation{Experimental Physics V, Center for Electronic
Correlations and Magnetism, Institute of Physics, University of Augsburg, 86135 Augsburg, Germany}

\author{A. K. Bera}
\email{anup.bera@helmholtz-berlin.de}
\affiliation{Helmholtz-Zentrum Berlin f\"{u}r Materialien und
Energie, 14109 Berlin, Germany}

\author{A. T. M. N. Islam}
\affiliation{Helmholtz-Zentrum Berlin f\"{u}r Materialien und
Energie, 14109 Berlin, Germany}

\author{B. Lake}
\affiliation{Helmholtz-Zentrum Berlin f\"{u}r Materialien und
Energie, 14109 Berlin, Germany}
\affiliation{Institut f\"{u}r Festk\"{o}rperphysik, Technische
Universit\"{a}t Berlin, 10623 Berlin, Germany}

\author{A.~Loidl}
\affiliation{Experimental Physics V, Center for Electronic
Correlations and Magnetism, Institute of Physics, University of Augsburg, 86135 Augsburg, Germany}

\author{J.~Deisenhofer}
\affiliation{Experimental Physics V, Center for Electronic
Correlations and Magnetism, Institute of Physics, University of Augsburg, 86135 Augsburg, Germany}

\date{\today}

\begin{abstract}
For quasi-one dimensional quantum spin systems theory predicts the
occurrence of a confinement of spinon excitation due to interchain
couplings.
Here we investigate the system \SrCoVO, a realization of the
weakly-coupled Ising-like XXZ antiferromagnetic chains, by terahertz
spectroscopy with and without applied magnetic field.
At low temperatures a series of excitations is observed, which split in a
Zeeman-like fashion in an applied magnetic field. These magnetic excitations
are identified as the theoretically predicted spinon-pair excitations. Using
a one dimensional Schr\"{o}dinger equation with a linear confinement potential
imposed by weak interchain couplings, the hierarchy of the confined spinons
can be fully described.
\end{abstract}

\pacs{75.30.-m,75.10.Pq,71.70.Ej}

\maketitle

\begin{figure}[t]
\centering
\includegraphics[width=80mm,clip]{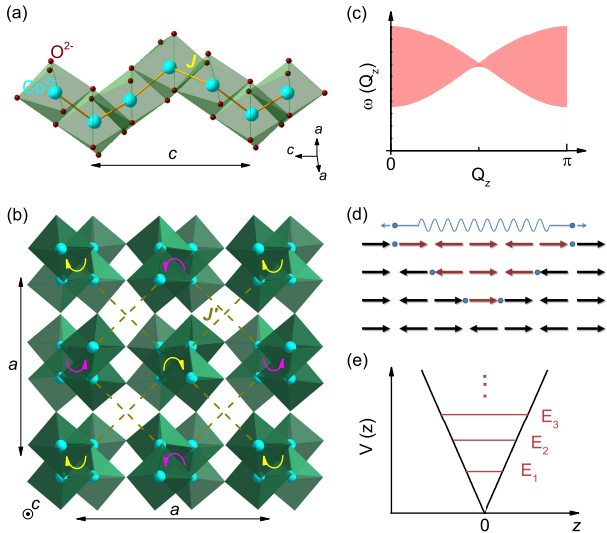}
\vspace{2mm} \caption[]{\label{Fig:Confinement}
(a) Screw chain consisting of CoO$_6$ octahedra running along the crystalline \emph{c} axis in \SrCoVO.
(b) Projection of left- and right-handed screw chains in the tetragonal \emph{ab} plane.
Interchain exchange interaction is dominated by
the antiferromagnetic interaction $J'$ between the nearest-neighbor Co ions in the \emph{ab} plane, which are from the neighboring chains with the same chirality.
(c) Shaded area indicates the excitation continuum of the spinon-pair quasiparticles corresponding to the Ising-like XXZ Hamiltonian
in Eq.~(\ref{Eq:Ising-Heisenberg}).\cite{Bougourzi98}
The spinon-pairs follow a quadratic dispersion relation along the chain direction close to the reciprocal $\Gamma$ point ($Q_z=0$).\cite{Shiba80a,Bougourzi98}
(d) Magnetic excitations of spinon-pairs with total pseudospin $\tilde{S} = 1$
corresponding to an odd number of pseudospin-1/2 flips.
The collinear antiferromagnetic order along the $c$ axis is stabilized below 5~K.\cite{Bera14}
(e) Quantized spinon-pair excitation levels due to linear confinement imposed by interchain exchange interaction.}
\end{figure}

Confinement of quasiparticle, a phenomenon mostly know from particle physics,\cite{Muta87}
can be also realized in condensed matter.\cite{Lake10,Coldea10}
Due to the strong interaction, quarks in a meson are asymptotically free at shorter distance and never exist as individual particles.
The counterpart in condensed matter, magnetic quasiparticle (spinon), exhibits similar confinement behaviors.
The spinon confinement can be illustrated in a paradigmatic one-dimensional system, Ising-like XXZ spin-1/2 chain,
where detailed theoretical descriptions have been achieved.
In a single Ising chain, two spinons (domain walls) are created by a spin flip.
The spinons, each carrying spin-1/2, can move freely along the chain by subsequent spin flips without cost of energy.
This results in a highly degenerate first excited state with energy $J$, the intrachain nearest-neighbor exchange interaction.
A finite transverse component of the exchange interaction lifts the degeneracy of
the first excited states, leading to an excitation continuum composed of two spinons
which propagate independently.\cite{Shiba80a} In the presence of weak interchain interactions,
two spinons are bound because the separation between them will frustrate interchain interactions [Fig.~\ref{Fig:Confinement}(b)-(d)].
The interchain interaction plays the role of an attractive potential,
proportional to the distance, confining the spinons into pairs. This leads to a
quantization of the excitation continuum into discrete levels of spinon bound states [Fig.~\ref{Fig:Confinement}(e)].

In this letter, we demonstrate experimentally such spinon confinement in the
weakly coupled spin-1/2 Ising-like antiferromagnetic chain \SrCoVO~by
high resolution terahertz spectroscopy.
\SrCoVO~crystalizes in a tetragonal structure with the space group $I4_1cd$.
The screw spin chains in \SrCoVO, consisting of edge-sharing CoO$_6$ octahedra,
propagate along the crystalline \emph{c} axis, as shown in the Fig.~\ref{Fig:Confinement}(a) and Fig.~\ref{Fig:Confinement}(b).\cite{Bera14,He2006}
Weak interchain couplings stabilize long-range collinear antiferromagnetic order below $T_N=5$~K.\cite{Bera14}
Spin-orbit coupling entangles the orbital and spin degrees of freedom of Co$^{2+}$,
making the total angular momentum a conserved quantity.
In the distorted octahedral environment, crystal field splitting is very strong that
the lowest-lying Kramers doublet of Co$^{2+}$ is well separated from the higher-lying spin-orbit quartet and sextet,
thus the magnetic moment of Co$^{2+}$ can be described by
a highly anisotropic total angular moment with \emph{pseudospin} $\tilde{S}=1/2$.\cite{Abragam1970}
Significant Ising-like anisotropy along the \emph{c} axis [Fig.~\ref{Fig:Confinement}(d)]
is also achieved by cooperative effects of the crystal field splitting and spin-orbit coupling.

Intrachain exchange interactions between the pseudospins in \SrCoVO~can be modeled by the Ising-like XXZ Hamiltonian\cite{Bonner64}
\begin{equation}\label{Eq:Ising-Heisenberg}
    J\sum_i[\tilde{S}^z_i\tilde{S}^z_{i+1}+\epsilon(\tilde{S}^x_i\tilde{S}^x_{i+1}+\tilde{S}^y_i\tilde{S}^y_{i+1})],
\end{equation}
where $J>0$ is the nearest-neighbor antiferromagnetic exchange interaction, and $0<\epsilon<1$ takes into account the Ising-like anisotropy arising from distortions of the CoO$_6$ octahedra.

High-quality single crystals of \SrCoVO~were grown in Helmholtz-Zentrum Berlin (HZB) using the floating-zone
method as described in Ref.~\cite{Bera14}.
Time-domain terahertz (THz) transmission measurements were performed
in the spectral range $0.1-3.3$~THz using a TPS Spectra 3000 spectrometer (TeraView Ltd.).
References (empty apertures) and samples were measured as function of temperatures.
Power spectra were obtained by doing Fourier transformation of the time-domain signals.
Magnetic field dependent transmission experiments were performed in Voigt geometry
with Backward Wave Oscillators covering
frequencies from 0.32 to 1.4 THz and a magneto-optical
cryostat (Oxford Instruments/Spectromag) with applied
magnetic fields up to 3~T.
Single crystals with typical size $4\times4\times0.5$~mm$^3$
were aligned with respect to the crystalline $a$ and $c$ axes.

\begin{figure}[t]
\centering
\includegraphics[width=85mm,clip]{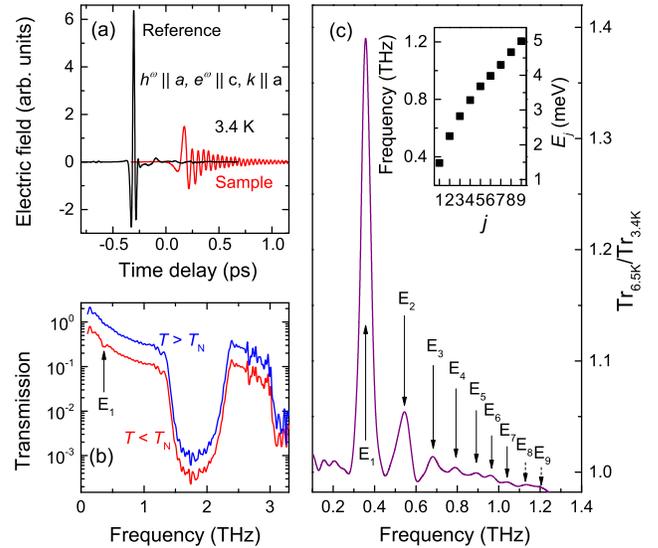}
\vspace{2mm} \caption[]{\label{Fig:TDS} (a) THz time domain signals of a reference and a \SrCoVO~sample measured at 3.4~K
with the electric field $e^\omega \parallel c$, the magnetic field $h^\omega \parallel c$, and wavevector $k\parallel a$.
(b) Transmission spectrum below and above $T_N$ in the THz spectral range
obtained from Fourier transformation of the time domain signal.
The spectrum measure above $T_N$ is shifted upward for clarity.
A dip corresponding to $E_1$ is marked in the spectrum measured below $T_N$.
The spectral range from 1.3 to 2.3 THz and above 3 THz is dominated by broad phonon bands.
(c) Ratio of transmission as obtained at 6.5~K and 3.4~K as a function of frequency below the phonon band.
Nine peaks can be observed and are indexed from $E_1$ to $E_9$.
Excitation energy $E_j$ versus energy level index $j=1,2,3,...$ are shown in the inset.}
\end{figure}

In Fig.~\ref{Fig:TDS}(a) we show
the electric-field amplitude of the transmitted electromagnetic wave through a reference and a \SrCoVO~sample
as a function of time delay in the magnetically ordered phase at 3.4~K.
The incident light propagating along the crystalline \emph{a} axis ($k$~$\|$~$a$) is polarized
with the \emph{ac} electric field parallel to the \emph{c} axis ($e^\omega$~$\|$~$c$) and the \emph{ac} magnetic field parallel to the \emph{a} axis ($h^\omega$~$\|$~$c$).
The transmission spectrum in the frequency domain [Fig.~\ref{Fig:TDS}(b)] is obtained by Fourier transformation of the time-domain signal.
From the power spectrum measured below $T_N$, one can notice an absorption line at about 0.35~THz marked as $E_1$.
This absorption line is not observed above $T_N$.
Figure~\ref{Fig:TDS}(c) shows the power spectrum at 6.5~K right above $T_N$ divided by the one measured at 3.4~K.
One can readily observe the sharp line $E_1$ followed by a series of absorption lines $E_2$,$E_3$,...,$E_9$ with increasing energies.
These absorption lines are not observed when the \emph{ac} magnetic field is parallel to the spin direction ($h^\omega$~$\|$~$c$).
This points to the magnetic nature of the absorption modes.
These modes can be assigned to magnetic excitations also based on their dependence on external magnetic field as discussed in the following.

\begin{figure*}[t]
\centering
\includegraphics[width=175mm,clip]{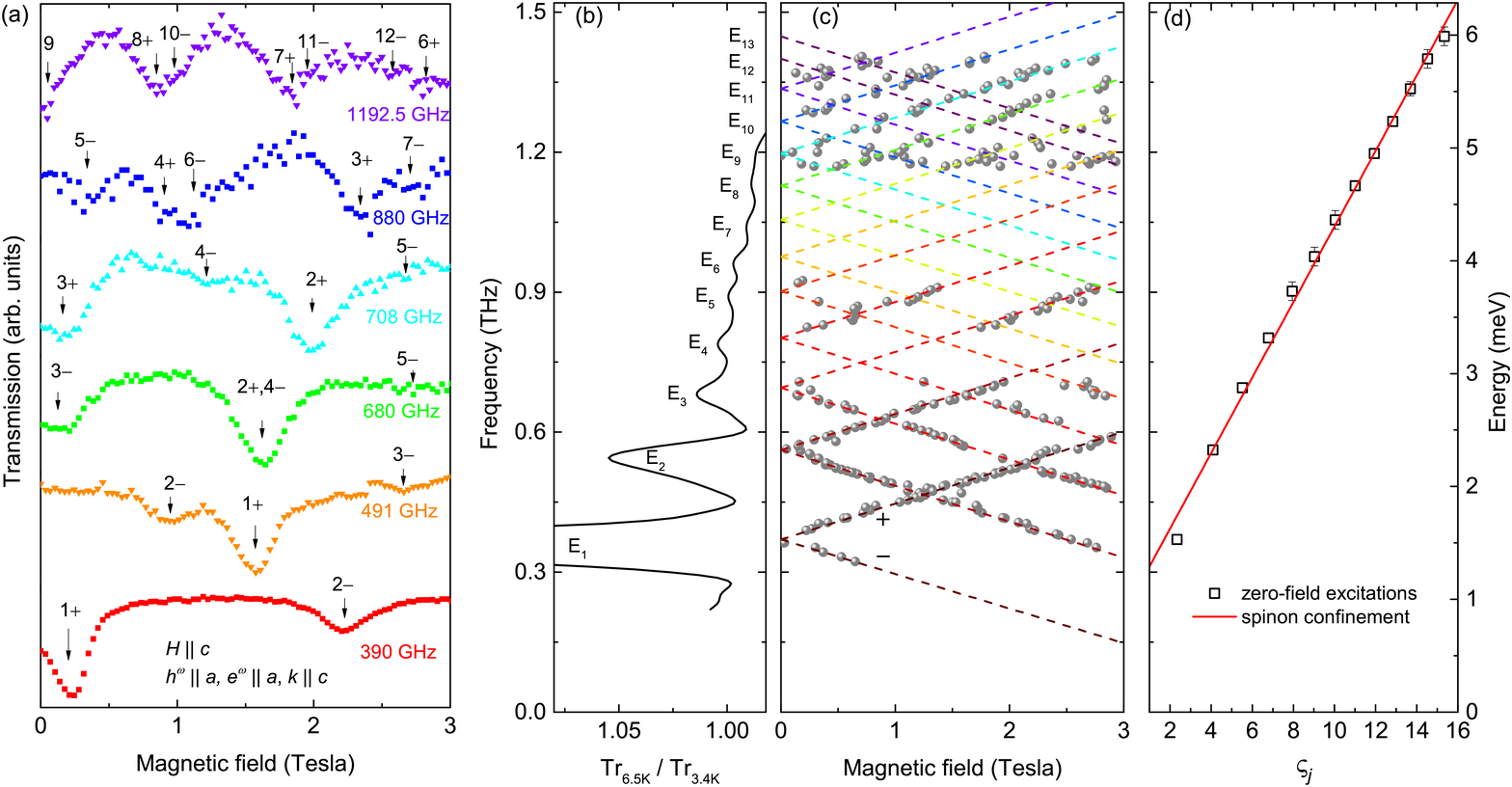}
\vspace{2mm} \caption[]{\label{Fig:HochFeldAnregungen} (a) Transmission spectra of various frequencies as a function of a longitudinal magnetic field at 2~K. Absorption modes are marked by the corresponding number of energy level in Fig.(b). (b) Normalized transmission spectrum at zero field adapted from Fig.~\ref{Fig:TDS}(c). (c) Energies of absorption modes as a function of applied longitudinal magnetic field. Their field dependence can be described by linear Zeeman term with a single \emph{g}-factor $g_\parallel=5.5$ (see text). (d) Hierarchy of the excitation energies at zero field can be modeled by the one dimensional Schr\"{o}dinger equation with linear confinement Eq.~(\ref{Eq:SpinnonSchroedingerEq}). $\zeta_j$s are the negative zeros of the Airy function (see text).}
\end{figure*}

The transmission spectrum of monochromatic THz light is measured at 2~K as a function of applied magnetic field along the chain direction,
which is also the direction of the Ising-like spins in the magnetically ordered phase.\cite{Bera14}
Typical spectra are shown in Fig.~\ref{Fig:HochFeldAnregungen}(a).
Several absorption modes can be observed at the resonance fields in the spectrum of each frequency.
The resonance magnetic fields shift with frequencies of the electromagnetic waves.
The absorption frequencies versus corresponding resonance magnetic fields are plotted in Figure~\ref{Fig:HochFeldAnregungen}(c).
One can clearly recognize a series of absorption modes, which
are split in magnetic field, with energies increasing or decreasing with increasing magnetic field.
All the lines follow a similar linear field dependence that can be described
by a Zeeman interaction term $\pm g_\parallel \mu_B H \tilde{S}$ with a single value
for the \emph{g}-factor $g_\parallel=5.5$, $\mu_B$ being the Bohr magneton, and $H$ the applied longitudinal magnetic field.
The total pseudospin $\tilde{S} = 1$ has been used due to the fact that two spinons each with $\tilde{S} = 1/2$ are involved.
Plus or minus signs correspond to split modes in magnetic field, and are also used to label the modes in Fig.~\ref{Fig:HochFeldAnregungen}(a) and Fig.~\ref{Fig:HochFeldAnregungen}(c).
The zero-field energies obtained by extrapolating the linear field dependence to zero field are shown in Fig.~\ref{Fig:HochFeldAnregungen}(d).
The zero-field spectrum of Fig.~\ref{Fig:TDS}(c) is shown in Fig.~\ref{Fig:HochFeldAnregungen}(b) for comparison.
The agreement in the energies for all observed excitations is clearly documented.

The observed magnetic excitations below $T_N$ exhibit clear hierarchical features:
The intensity of the lines decreases with increasing energy
and the energy difference between neighboring excitations also decreases with increasing energy [see Fig.~\ref{Fig:TDS}(c) and inset].
These are exactly the features predicted by theory for spinon bound states due to confinement.\cite{Shiba80}
The excitations are indeed the spinon-pair excitations with total pseudospin $\tilde{S}=1$ as confirmed by their field dependence.
Every spinon-pair excitation is doubly degenerate in zero field,
and the degeneracy is lifted in a finite longitudinal external magnetic field [Fig.~\ref{Fig:HochFeldAnregungen}(b)(c)].
Energies of the non-degenerate spinon-pair excitations increase or decrease depending on whether the total pseudospin is parallel or antiparallel to the external field, respectively.

The $\tilde{S} = 1$ spinon-pair bound-states correspond to an odd number of pseudospin flips between the spinons as illustrated in Fig.~\ref{Fig:Confinement}(d).\cite{Shiba80a,Goff95}
Such bound states have also been observed in the related compound \BaCoVO.\cite{Kimura07}
The other series of spinon-pair excitations with $\tilde{S}=0$ observed in \BaCoVO~by neutron scattering experiments \cite{Grenier14}
are not observed in these THz measurements [see Fig.~\ref{Fig:HochFeldAnregungen}(c)].
This confirms the magnetic dipole active nature of the $\tilde{S} = 1$ spinon-pair excitations.
For a spinon pair, the relative motion of one spinon with respect to the other along the chain can be described by the one-dimensional Schr\"{o}dinger equation
\begin{equation}\label{Eq:SpinnonSchroedingerEq}
    -\frac{\hbar^2}{\mu}\frac{d^2\varphi}{dz^2}+\lambda |z| \varphi=(E-2E_0)\varphi
\end{equation}
with the linear potential $\lambda |z|$ taking into account the confinement imposed by the interchain antiferromagnetic exchange interaction $J'$ [Fig.~\ref{Fig:Confinement}(b)], i.e., $\lambda=2J'\langle \tilde{S}_z\rangle^2/c$.
\emph{c} is the lattice constant along the chain direction [Fig.~\ref{Fig:Confinement}(a)].
A quadratic dispersion relation at the $\Gamma$ point has been assumed,
following the calculations of the two-spinon dynamic structure factor [Fig.~\ref{Fig:Confinement}(b)].\cite{Bougourzi98}
Since the photons have a linear dispersion relation with the speed of light involved,
THz spectroscopy probes only the reciprocal space that is very close to the $\Gamma$ point.

The solution of Eq.~(\ref{Eq:SpinnonSchroedingerEq}) gives the eigenenergies of the spinon-pair bound states\cite{McCoy78}
\begin{equation}\label{Eq:SpinnonSchroedingerSolution}
    E_j=2E_0+\zeta_j\lambda^{2/3}\left(\frac{\hbar^2}{\mu}\right)^{1/3} \hspace{5mm}  j=1,2,3,...
\end{equation}
The sequence of the bound states is specified by the prefactors $\zeta_j$ that are the negative zeros of the Airy function $Ai(-\zeta_j)=0$, $\zeta_j=2.338, 4.088, 5.520, ...$ ($j=1,2,3,...$).
$2E_0$ is the energy threshold for creating a free spinon pair.
The bound-state energies at zero field exhibit a linear dependence on $\zeta_j$ as shown in Fig.~\ref{Fig:HochFeldAnregungen}(d).
The solid line in Fig.~\ref{Fig:HochFeldAnregungen}(d) is obtained by fitting the experimental results to Eq.~(\ref{Eq:SpinnonSchroedingerSolution}).
One can see that the energy hierarchy of the spinon-pair bound-states is nicely modeled with $2E_0=0.96$~meV and $\lambda^{2/3}(\frac{\hbar^2}{\mu})^{1/3}=0.33$~meV.

The energy threshold $2E_0$
is significantly smaller than the pure Ising-limit $J=60$~K~$\simeq 5.17$~meV.
The value has been obtained by fitting the magnetic susceptibility using a
pure Ising model.\cite{Bera14,Bonner64}
The fact that $2E_0\ll J$ stems from the finite $\epsilon$ term in Eq.~(\ref{Eq:Ising-Heisenberg}).\cite{Shiba80a,Bougourzi98}
The deviation of $\epsilon$ from 1 is a measure of the anisotropy arising from distortions of the CoO$_6$ octahedra,
which is also reflected by the $g_\parallel$ factor.
If we assume a trigonal distortion with strength $\delta$,\cite{Abragam51,Lines63}
then the induced anisotropy depends on the renormalized parameter $\delta/\Lambda$
with $\Lambda<0$ being the spin-orbit coupling.
The obtained value of $g_\parallel=5.5$ corresponds to $\delta/\Lambda=-0.80$ and $\epsilon=0.73$.\cite{Lines63}
A smaller value of $\epsilon=0.41$ in \BaCoVO~indicates a larger distortion of the CoO$_6$ octahedra.\cite{Lines63,Grenier14}
This is also indicated by a larger value of $g_\parallel=6.2$ and
a larger excitation gap of 1.7~meV in \BaCoVO~with a similar intrachain exchange interaction.\cite{Kimura07,Lines63,Shiba80a,He2005,Bera14}

Spinon bound states in the quasi-one-dimensional antiferromagnets \SrCoVO~and \BaCoVO~can
be classified by their pseudospins which are either $\tilde{S}=1$ or $\tilde{S}=0$,
analogous to the classification of mesons (quark-antiquark bound states) by their isospins (either 1 or 0).
This is in clear distinction to the quasi-one-dimensional ferromagnet CoNb$_2$O$_6$ where pseudospin is not a conserved quantity,
since more spins are flipped at higher energies.
Even at a certain energy level, spin blocks (domains) with different lengths in CoNb$_2$O$_6$ tend to flip with comparable probablities.\cite{Morris14}


In the strong interaction scenario, the strings between quarks can snap and
heavy hadron particles are expected to decay into lighter ones,
when creating heavier particles is energetically less favorable.
Analogous to this scenario, spinon bound states can be expected only up to an energy of $E_1 + E_1$,
above which an energy continuum would appear, as reported in CoNb$_2$O$_6$.\cite{Morris14}
This continuum is surprisingly not observed in \SrCoVO~or in \BaCoVO~up to the energy much higher than $E_1 + E_1$.
The difference is possibly due to the different strengths of spinon-spinon interactions
in the two systems.
The corresponding parameter $\lambda/\sqrt{\mu}$ in \SrCoVO~is
larger by one order of magnitude than in CoNb$_2$O$_6$.
The $E_1 + E_1$ states with pseudospin $\tilde{S}=2$ and
any other bound states created by higher-order confinement processes with larger pseudospins,
such as bound states of $E_1$ states in neighboring chains,\cite{Morris14}
can be excluded in \SrCoVO~as documented in Fig.~\ref{Fig:HochFeldAnregungen}(c),
because a stronger field dependence expected for these states is absent.

Using THz transmission spectroscopy also in external magnetic fields, we have investigated low-energy magnetic excitations in the quasi-one-dimensional Ising-like antiferromagnet \SrCoVO.
Spinon-pair excitations on the antiferromagnetic ground state have been observed in this XXZ-type antiferromagnet.
Spinon-pair bound states with entangled spin-orbit moment $\tilde{S}=1$ are determined unambiguously.
The hierarchy of the spinon-pair bound-states can be described by a one-dimensional Schr\"{o}dinger equation
with a linear confinement potential imposed by the interchain interaction.
Energy continuum of the elementary spinon-pair excitations is surprisingly not observed in \SrCoVO.

We acknowledge helpful discussion with N. Peter Armitage.
Partial support by the Deutsche Forschungsgemeinschaft via TRR 80
(Augsburg - Munich - Stuttgart) and by the Project DE 1762/2-1 is acknowledged.


\begin{thebibliography}{39}

\bibitem{Muta87} T. Muta, \emph{Foundations of Quantum Chromodynamics}, (World Scientific Publishing, Singapore, 1987).

\bibitem{Lake10} B. Lake, A. M. Tsvelik, S. Notbohm, D. A. Tennant, T. G. Perring, M. Reehuis, Ch. Sekar, G. Krabbes, and B. B\"{u}chner, Nature Physics \textbf{6}, 50 (2010).


\bibitem{Coldea10} R. Coldea, D. A. Tennant, E. M. Wheeler, E. Wawrzynska,
D. Prabhakaran, M. Telling, K. Habicht, P. Smeibidl, and
K. Kiefer, Science \textbf{327}, 177 (2010).


\bibitem{Shiba80a} N. Ishimura and H. Shiba, Prog. Theor. Phys. \textbf{63}, 743 (1980).

\bibitem{Bera14} A. K. Bera, B. Lake, W.-D. Stein, and S. Zander, Phys. Rev. B \textbf{89}, 094402 (2014).

\bibitem{He2006} Z. He, T. Taniyama, and M. Itoh, Phys. Rev. B \textbf{73}, 212406 (2006).

\bibitem{Abragam1970} A. Abragam and B. Bleaney, {\it Electron
Paramagnetic Resonance of Transition Ions}, (Clarendon, Oxford,
1970).

\bibitem{Bonner64} J. C. Bonner and M. E. Fisher, Phys. Rev. \textbf{135}, A640 (1964).


\bibitem{Bougourzi98} A. H. Bougourzi, M. Karbach, and G. M\"{u}ller, Phys. Rev. B \textbf{57}, 11429 (1998).


\bibitem{Shiba80} H. Shiba, Prog. Theor. Phys. \textbf{64}, 466 (1980).

\bibitem{Goff95} J. P. Goff, D. A. Tennant, and S. E. Nagler, Phys. Rev. B \textbf{52}, 15992 (1995).


\bibitem{Kimura07} S. Kimura, H. Yashiro, K. Okunishi, M. Hagiwara, Z. He, K. Kindo, T. Taniyama, and M. Itoh, Phys. Rev. Lett. \textbf{99}, 087602 (2007).

\bibitem{Grenier14} B. Grenier, S. Petit, V. Simonet, L.-P. Regnault, E. Can\'{e}vet, S. Raymond, B. Canals, C. Berthier, and P. Lejay, arXiv:1407.0213v1 (unpublished).


\bibitem{McCoy78} B. M. McCoy and T. T. Wu, Phys. Rev. D \textbf{18}, 1259 (1978).


\bibitem{Lines63} M. E. Lines, Phys. Rev. \textbf{131}, 546 (1963).

\bibitem{Abragam51} A. Abragam and M. H. L. Pryce, Proc. R. Soc. A \textbf{206}, 173 (1951).


\bibitem{He2005} Z. He, T. Taniyama, T. Ky\^{o}men, and M. Itoh, Phys. Rev. B \textbf{72}, 172403 (2005).

\bibitem{Morris14} C. M. Morris, R. Vald\'{e}s Aguilar, A. Ghosh, S. M. Koohpayeh,
J. Krizan, R. J. Cava, O. Tchernyshyov, T. M. McQueen, and N. P. Armitage, Phys. Rev. Lett. \textbf{112}, 137403
(2014).


\end{thebibliography}
\end{document}